\documentclass[a4paper,10pt]{article}

\usepackage{amsmath}
\usepackage{amsfonts}
\usepackage{amssymb}
\usepackage{array}
\usepackage{multirow}
\usepackage{hyperref}
\usepackage[all]{hypcap}
\usepackage{graphicx} 

\newcommand{\bea}{\begin{eqnarray}}
\newcommand{\eea}{\end{eqnarray}}

\begin{document}

\title{The Physics Role and Potential of future Atmospheric Detectors\footnote{Based on a Plenary Talk at NUFACT 09, Chicago} }


\author{Raj Gandhi\\ \small{\textit{Harish-Chandra Research Institute, Jhunsi, Allahabad, India-211019}}}



\date{}
\maketitle

\begin{abstract}
We discuss the physics capabilities of basic types of future atmospheric detectors being considered at present, with  their 
 strengths and limitations, and compare them with those of long baseline (LBL) experiments. We also argue  that recent studies signal the importance of synergistically combining complementary features of both these classes of experiments in order to accrue maximum benefit towards furthering our goal of 
building a complete picture of neutrino properties and parameters. 

\end{abstract}


\section{Introduction}

Atmospheric neutrinos, via the experiments observing them, have served the cause of particle physics well. The initial observations of $\nu_\mu, \nu_e$ fluxes by the iron calorimeters Frejus \cite{fre} and NUSEX \cite{nus} and their deficits, first reported by the early  Water Cerenkov detectors IMB \cite{imb} and Kamiokande \cite{kam} set the stage for intensive subsequent experimental activity. This culminated in 1998, with Super-Kamiokande \cite{sk} reporting a high statistics 
result for $\nu_\mu\rightarrow\nu_\tau$ oscillations, providing the first definitive evidence for neutrino mass and physics beyond the Standard Model. These findings were   reconfirmed  by the subsequent calorimetric measurements of SOUDAN-2 \cite{Allison:1999ms} and MACRO \cite{Ambrosio:2001je}. More recently, their accuracy has been buttressed by complementary measurements of the atmospheric oscillation parameters $\Delta m^2_{atm}$ and the mixing angle $\theta_{23}$ by the accelerator experiments K2K \cite{k2k} and MINOS \cite{min}.

This impressive history notwithstanding, recent developments and the present state of neutrino physics warrant a realignment of the role of current and planned atmospheric neutrino experiments vis a vis long baseline projects. In particular, this is required by the shift in the priorities of the field itself: unlike the past, precision measurements of neutrino mass and mixing parameters must be made in tandem with the earlier (and still currently very important)  goal of discovering new physics. The fact that these two objectives are inter-dependant imposes special requirements and constraints which must be taken into consideration.

In this note, we review the physics capabilities of basic types of future atmospheric detectors being considered at present, their 
 strengths and limitations, and compare them with those of long baseline (LBL) experiments. We also argue  that recent studies signal the importance of synergistically combining complementary features of both these classes of experiments in order to accrue maximum benefit towards furthering our goal of 
building a complete picture of neutrino properties and parameters. 

\section{Atmospheric Neutrino Detectors vs LBL Experiments: A Qualitative Comparison}

Despite their stellar role as discovery tools in the recent past, atmospheric detectors have certain inherent limitations that constrain their performance as instruments capable  of precision measurements. The absolute values of atmospheric fluxes of $\nu_\mu$ and $\nu_e$ are uncertain by as much as $10-20\%$, although their ratios are much better known. These fluxes are sharply dropping functions of energy, with $\frac{d\phi_\nu}{dE}\sim E^{-\gamma}$ and $\gamma \simeq 3$ for $\nu_\mu$ and $\simeq 3.5$ for 
$\nu_e$. At energies where fluxes are significant ($\it{i.e.} \leq 2-3$ GeV), the total neutrino-nucleon cross-section is dominated by quasi-elastic contributions which have significant uncertainties. Addtionally, Fermi motion of nucleons renders accurate energy reconstruction of the incoming neutrino difficult.  Moreover, the produced charged lepton direction and that of the incoming neutrino can be significantly different at low energies. The upshot of all of these factors is a lack of precision in 
an event by event determination of $L$ and $E$. Consequently, determinations of $\Delta m^2$ (which depend on $L/E)$ by atmospheric experiments cannot match the precision that LBL setups can achieve with much smaller data sets.
 Similarly, the precision expected to be achieved by experments like T2K \cite{t2k} and NO$\nu$A \cite{nova} for $sin^22\theta_{23}$ is also expected to be correspondingly higher than what may be possible in future atmospheric detectors. 

A distinct advantage of atmospheric detectors, however,  is the broad band in both $L$ (20 km to 12500 km) and $E$ (100 MeV to 10 TeV) that they can tap into. This infuses an intrinsic  complementarity into their physics capabilities compared to the fixed $L$ and relatively narrow band of $E$ present in LBL experiments. One of the main goals of this note is to emphasize the usefulness and necessity of pursuing an approach that combines, at the level of analysis, the benefits of both classes of experiments.

A feature which is expected to be common to all types of detectors  in the future is a large
fiducial volume, leading to the ability to accrue statistics at rates not possible for existing atmospheric experiments. This is essential for any contempleted combined operation and subsequent data analysis of such a detector with a LBL experiment.

 Among the problems  LBL experiments
must confront is the fact that  their results are beset with  several distinct types of parameter degeneracies \cite{cam57,cam58,cam59,cam60}
 which we describe briefly here:\\
a) The intrinsic, or $\left\lbrace \mathrm{\delta_{CP}}, \theta_{13} \right\rbrace$ degeneracy, which arises
when different pairs of values of the parameters
${\mathrm{\delta_{CP}}}$ and $\theta_{13}$ give the same neutrino
and anti-neutrino oscillation probabilities, assuming other
parameters to be known and fixed.
This may be expressed as 
\bea
{\mathrm{P_{\alpha\beta}({\mathrm{\delta_{CP}}}, \theta_{13})}} &=&
{\mathrm{P_{\alpha\beta}({\mathrm{\delta_{CP}'}}, \theta_{13}')}} \nonumber \\
{\mathrm{{\bar P}_{\alpha\beta}({\mathrm{\delta_{CP}}},
 \theta_{13})}}
&=&
{\mathrm{{\bar P}_{\alpha\beta}({\mathrm{\delta_{CP}'}}, \theta_{13}')}}
\label{eq:delcpth13}
\eea
b) The octant, or  $\theta_{23}, (\pi/2 -\theta_{23})$ degeneracy, which arises primarily due the LBL experiments being mainly sensitive to $\sin^22\theta_{23}$. As in a), one obtains two solutions of equal statisitical significance, but associated with different pairs of values of $({\mathrm{\delta_{CP}}}, \theta_{13})$.\\
c) Similarly, the {{mass hierarchy degeneracy}} occurs due to identical solutions
for P and ${\mathrm {\bar P}}$ for different pairs of ${\mathrm{\delta_{CP}}}$
and $\theta_{13}$
 with opposite signs of $\Delta_{31}$ (fixing other parameters):{
\bea
{\mathrm{P_{\alpha\beta}({\mathrm{\Delta_{31}>0}},{\mathrm{\delta_{CP}}}, \theta_{13})}}
={\mathrm{P_{\alpha\beta}({\mathrm{\Delta_{31}<0}},{\mathrm{\delta_{CP}'}}, \theta_{13}')}} \nonumber \\
{\mathrm{{\bar P}_{\alpha\beta}({\mathrm{\Delta_{31}>0}},{\mathrm{\delta_{CP}}},\theta_{13})}}
={\mathrm{{\bar P}_{\alpha\beta}({\mathrm{\Delta_{31}<0}},{\mathrm{\delta_{CP}'}}, \theta_{13}')}}
\label{eq:sgndelmsq}
\eea

Consequently, while providing high precision measurements of $|\Delta m^2_{31}|$ and $\sin^2 2\theta_{23}$, LBL experiments of the (near) future will not be able to provide 
definitive information on the mass hierarchy, the CP phase or the octant in which $\theta_{23}$ lies. After discussing the basic types of atmospheric detectors in various stages of planning and construction, we discuss how combining their capabilities with LBL experiments  maximizes the physics potential and offers an opportunity to resolve these questions. This occurs primarily due to their contributing  data from a large range of $E$ and $L$ (even though the precision is relatively lower) to complement the precision measurement carried out by an LBL setup at one fixed value of $L$ and a narrow band in $E$.

\section {Future Atmospheric Detectors}

We first review the basic types of such experiments, along with their proposed locations
and features.

{\bf Water Cerenkov Detectors:} With an impressive history of succesful physics contributions \cite{imb,kam,sk} this remains the best understood type of atmospheric detector. The detection medium is cheap and stable, making large volumes fiscally feasible, with most of the cost residing in the acquisition of photo-multipler tubes and purification systems. 

A Water Cerenkov detector provides a well-understood mode of detection and separation of $\mu$ and $e$ leptons via ring topology. It has a lower energy threshold compared
to a magnetized iron calorimeter. Future projects planned include\\ i) a large modular detector, (part of the DUSEL setup)   at the Homestake mine in South Dakota\cite{uno}
, with a proposed 300 kT fiducial mass and a 1300 km baseline originating at Fermilab, located at a depth of 4800 mwe (meters of water equivalent).\\
ii) A second major project is Hyper-Kamiokande (HK), in Tochibora, Japan \cite{hk}, with a 550 kT
fiducial mass, a 290 km baseline and a location depth of 1500 mwe.\\
iii) In Europe, MEMPHYS, located at Frejus \cite{mem}, at a distance of 130 km from CERN, has a proposed fiducial mass of 440 kT, at a depth identical to that of the detector at Homestake in i) above.   

{\bf Magnetized Iron Calorimeter Detectors:} The major planned project in this category of detector is INO, in Southern India \cite{ino}. It will be 50-100 kT in mass, located at about 4000
mwe, and it will be sensitive to muons only.The detection threshold will be $1-2$ GeV and its distance from a CERN or other Europeon beam would be in the neighbourhood of 
7000 km. Its charge identification capability gives it an edge for hierarchy determination and CP, CPT studies. The high Z
medium also allows a studies of very high energy cosmic-ray (CR) muons using the pair-meter method, allowing a probe of the CR flux at energies around the "knee" and beyond \cite{vhe}.

{\bf{Liquid Argon Detectors:}} These offer bubble-chamber like  imaging of ionization tracks of drift electrons, in addition to scintillation and  Cerenkov light readout. They also have a low energy threshold  for the  detection of leptons, as long as charge identification is not a requirement. The  inclusion of B field adds a charge identification capability for muons above 800 MeV and for electrons above 1 GeV.

In general, Liquid Argon as a medium, offers  superior particle identification {($e/\mu/\pi/p$ separation)} and calorimetry but the technology remains relatively untested compared to  the well-understood Water Cerenkov and Iron Calorimeter detectors. A proposed effort is GLACIER, a 100 kT Liquid Argon European experiment \cite{gla}. Also being planned is possible 100 kT Liq Ar detector for the DUSEL project in the US \cite{uno}.

\section{Physics Prospects}

This section discusses the role future atmospheric detectors can play towards the  achievement of the outstanding physics goals which guide the overall neutrino program today. We consider each of the major unresolved issues in turn, and discuss the progress possible via atmospheric experiments. We point out, on the basis of recent studies, that in several cases, the synergistic combination of an atmospheric detector with a beam experiment leads to results which cannot be achieved by either experiment alone.

{\bf Determination of $\theta_{13}$:} In principle, an atmospheric detector has sensitivity to $\theta_{13}$ via matter effects in the earth as manifested in event rate channels sensitive to both $P_{\nu_\mu\rightarrow\nu_e}$ and $P_{\nu_\mu\rightarrow\nu_\mu}$. A strong manifestation of these effects however, is confined to the band of $E= 4-10$ GeV, as shown for the muon survival probability in Figure 1 \cite{rg1,rg2}. The corresponding $L$ range where these effects appear, is about 4000-10000 km. 

\begin{figure}[t]
  \centering
  \includegraphics[width=.85\textwidth]{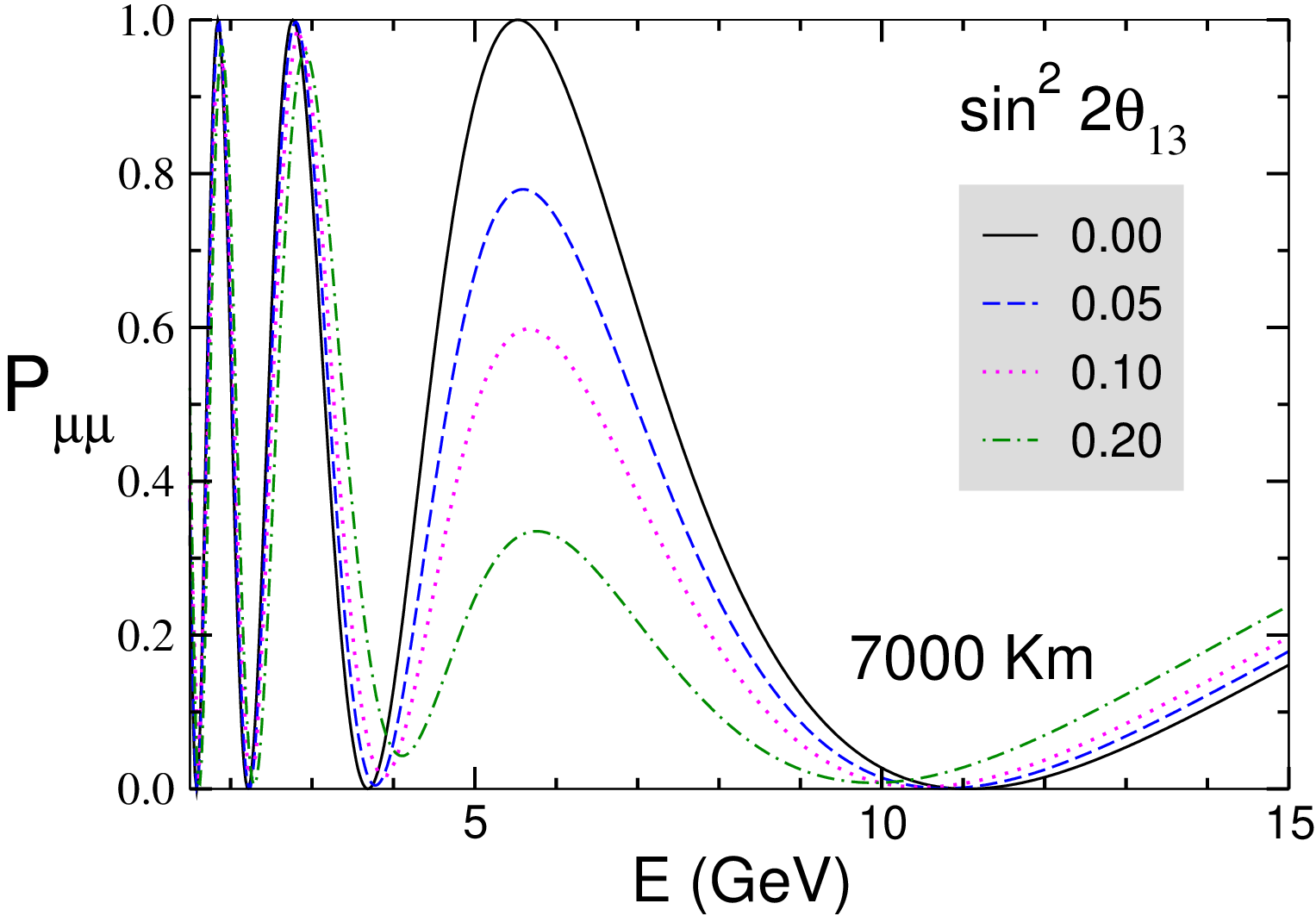}
  \caption{Sensitivity of the muon survival probability to $\sin^22\theta_{13}$ at 
very long baselines and energies of 4-10 GeV. From \cite{rg2}.} 
\end{figure}

\begin{figure}[h]
  \centering
  \includegraphics[width=.85\textwidth]{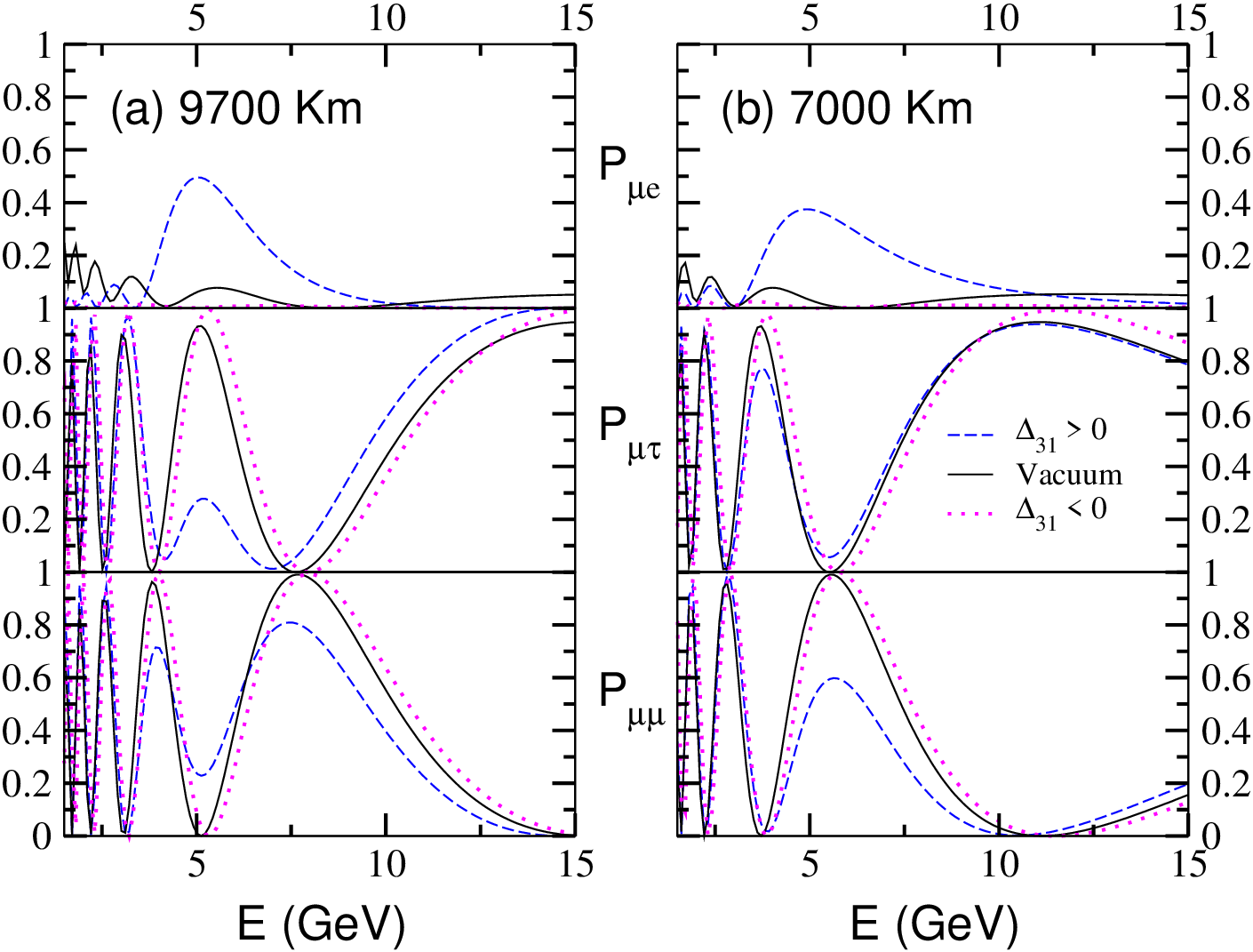}
\caption{The variations in $P_{\nu_\mu\rightarrow\nu_e}$, $P_{\nu_\mu\rightarrow\nu_\tau}$, $P_{\nu_\mu\rightarrow\nu_\mu}$ for both types of hierarchy (normal, NH) and inverted, IH) at long baselines. From \cite{rg2}.}
\end{figure}

In practice, event rates at these energies are suppressed by falling fluxes, and the effects, being matter based, appear either in the lepton or the antilepton sector, not both. For Water Cerenkov and Liquid Argon detectors, the $P_{\nu_\mu\rightarrow\nu_e}$
channel, while providing a handle not available to iron calorimeters, is compromised by the background from the $\nu_e\rightarrow\nu_e$ survival
probability.
When we take this into account with the fact that T2K, NO$\nu$A and the reactor experiments Double Chooz \cite{dch} and Daya Bay \cite{day} will soon provide more accurate measurements of $\theta_{13}$ on a significantly shorter timescale, it is apparent that atmospheric detectors will not have much of a role to play in accurate measurements of this parameter.

{\bf Determination of the Mass Hierarchy:}\
 For values of $sin^22\theta_{13}$ which are not too small, matter effects in atmospheric detectors  bring out features which may allow a determination of the hierarchy, albeit with about a decade of data taking in a large detector. In particular, an iron calorimeter like INO offers 
an advantage due to its capability to identify the muon charge. Figure 2 shows the variations in $P_{\nu_\mu\rightarrow\nu_e}$, $P_{\nu_\mu\rightarrow\nu_\tau}$, $P_{\nu_\mu\rightarrow\nu_\mu}$ for both types of hierarchy (normal, NH) and inverted, IH) at long baselines. It is also important to note that in a detector which relies solely on the muon survival channel for its data, like INO, the problem of the intrinsic, or {$({\mathrm{\delta_{CP}}}, \theta_{13})$ degeneracy is ameliorated \cite{rg2}. Figure 3, from \cite{rg3}   demonstrates the capability of such a detector for hierarchy determination over a 1 Mton-yr exposure.

\begin{figure}[t]
  \centering
  \includegraphics[width=.85\textwidth]{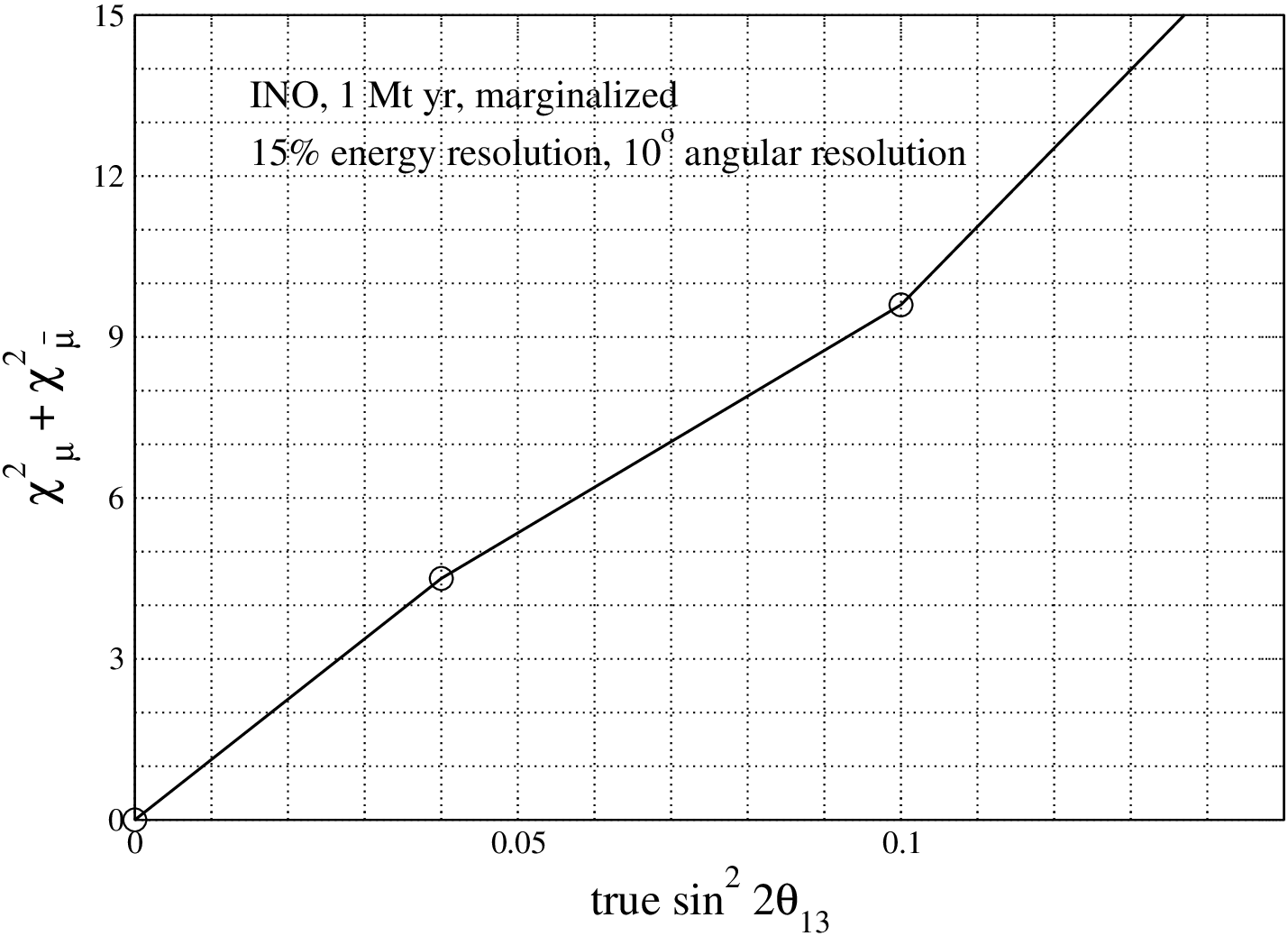}
  \caption{Sensitivity with respect to $\theta_{13}$ of a large magnetized Iron Calorimeter to the mass hierarchy for an exposure of 1 Mton-yr. From \cite{rg3}.}
\end{figure}

In Figure 4, from \cite{aut37}, we show how the combination of various LBL beam  experiments with atmospheric data collected by the proposed Water Cerenkov detector MEMPHYS greatly enhances the prospects of hierarchy determination. On their own, currently contemplated  LBL projects have baselines which do not allow matter effects to develop fully, thus their sensitivity to the hierarchy determination is very low.

\begin{figure}[h]
  \centering
  \includegraphics[width=.85\textwidth]{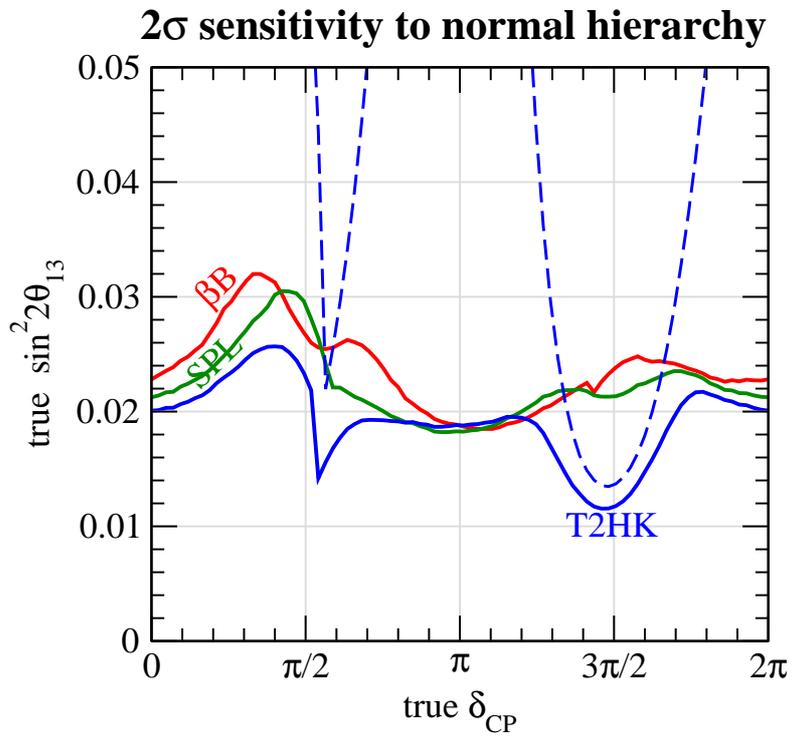}
\caption{Hierarchy sensitivity of CERN-MEMPHYS LBL on its own (dashed curves)
and when combined with atmospheric data (solid curves). From \cite{cam}.}
\end{figure}

{\bf Determining the octant of $\theta_{23}$:} Atmospheric experiments are sensitive to the deviation $\sin^2\theta_{23}$ from $1/2$ via an excess $\Delta n_e$ of electron events  detectable in the sub-GeV region  for Water Cerenkov and Liquid Argon detectors. This is given by \cite{per}
\begin{eqnarray}
\label{simple1}
 \Delta n_e &=& (1/2-\sin^2\theta_{23})\frac{\phi_{\mu}^0}{\phi_{e}^0}
P_{2f}(\Delta m^2_{21},\theta_{12})
\end{eqnarray} 
This excess at low energies depends on solar parameters, as is manifest above, and the dependance of  $\theta_{13}$ is  sub-dominant. A second channel that is sensitive to the octant is in the GeV region, and here the sensitivity manifests itself as a decrease in muon events. There is a strong dependance on $\theta_{13}$, as discussed in 
\cite{san}.

Figure 5, from \cite{hms} demonstrates the capabilities of LBL and atmospheric  detectors on their own
as octant discriminators, and the improvement that results when their capabilities are combined. One notes the fact that even though the octant sensitivity of LBL experiments on their own is negligble compared to that of an  atmospheric detector, the improved precision in the values of  parameters like $\sin^2\theta_{23}$ and $|\Delta m^2_{31}|$ which they bring to the  analysis helps achieve a significant improvement in sensitivity for the combination of the two.

\begin{figure}[t]
  \centering
  \includegraphics[width=.85\textwidth]{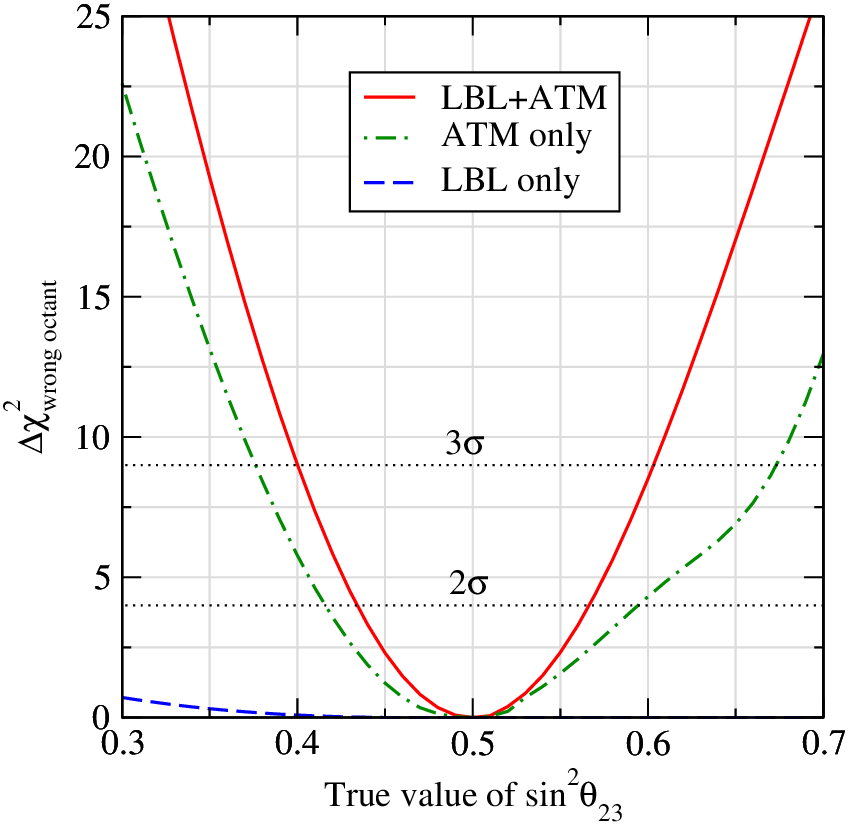}
  \caption{Octant discrimination capabilities of LBL and  atmospheric detectors, both individually and when combined. From \cite{hms}.}
\end{figure}

{\bf Atmospheric Detectors as degeneracy resolvers:} Finally, we demonstrate the power of future atmospheric detectors to lift the degeneracies that are inherent to beam experiments. Using the CERN-SPL beam as an example, and the MEMPHYS detector located at a baseline of 130 km, with a 440 kT fiducial mass, the left panel in Figure 6 \cite{cam} shows the full 8 fold degeneracy that was discussed above, appearing in SPL rates data
simulated over a period of 10 years, for both the apppearance and disappearance channels. The right panel of the same figure demonstrates, via the solid curves, how the addition of spectral information from the two channels reduces this to a four-fold degeneracy. Full resolution, and identification of the true solution is, however, only achieved when atmospheric data are included (for the same detector), as shown by the solid red region.

\begin{figure}[h]
  \centering
  \includegraphics[width=0.95\textwidth]{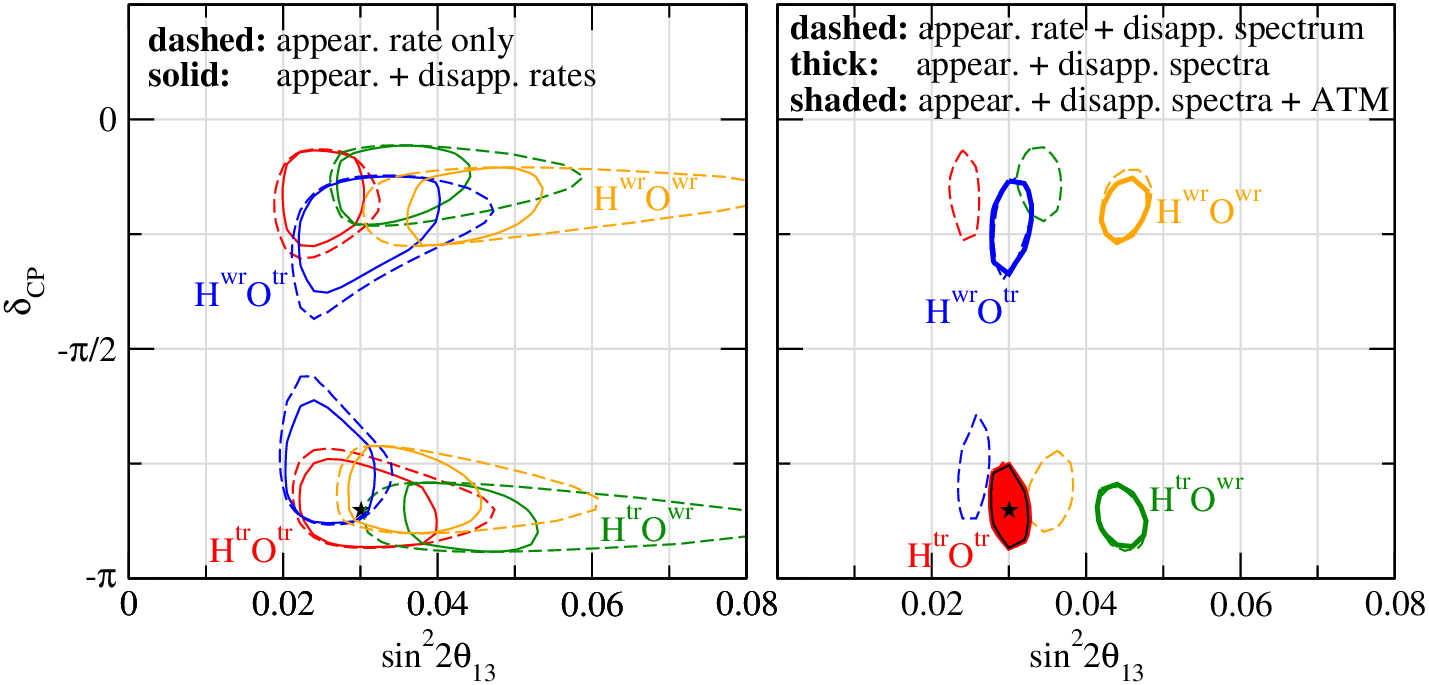}
  \caption{8 fold Degeneracy resolution by combining LBL and atmospheric data, as described in the text. From \cite{cam}.}
\end{figure}

\section{Summary and Conclusions:} We have entered  an era where the goals of neutrino physics demand both precision measurements of known parameters  and dicovery oriented experiments. The requirements imposed by these somewhat tangential goals can be difficult to achieve in any one given class of experiment. In this note we have tried to show how combining the positive features of large  mass atmospheric experiments, with 
their access to a wide range  of $L$ and $E$ values, with data garnered by the upcoming  precision oriented LBL projects can lead to a fruitful physics program that seeks to meet both these goals over the next decade.


\clearpage

\begin{thebibliography}{9}

\bibitem{fre}
{\bf Fréjus} Collaboration, K.~Daum {\em et.~al.},  {\em Z. Phys.} {\bf
  C66} (1995) 417--428.
\bibitem{nus}
{\bf The NUSEX} Collaboration, M.~Aglietta {\em et.~al.},  {\em Europhys. Lett.}
  {\bf 8} (1989) 611--614.
\bibitem{imb}
R.~Becker-Szendy {\em et.~al.}, {\em Phys. Rev.} {\bf D46} (1992)
  3720--3724.
\bibitem{kam}
{\bf Kamiokande-II} Collaboration, K.~S. Hirata {\em et.~al.}, {\em Phys.
  Lett.} {\bf B280} (1992) 146--152.
\bibitem{sk}
{\bf Super-Kamiokande} Collaboration, Y.~Fukuda {\em et.~al.},  {\em Phys. Rev. Lett.} {\bf 81} (1998)
  1562--1567
\bibitem{Allison:1999ms}
{\bf Soudan-2} Collaboration, W.~W.~M. Allison {\em et.~al.}, ``The atmospheric
  neutrino flavor ratio from a 3.9 fiducial kiloton-year exposure of Soudan 2''
  {\em Phys. Lett.} {\bf B449} (1999) 137--144
\bibitem{Ambrosio:2001je}
{\bf MACRO} Collaboration, M.~Ambrosio {\em et.~al.},  {\em Phys. Lett.} {\bf
  B517} (2001) 59--66 \href{http://arXiv.org/abs/hep-ex/0106049}{{\tt
  hep-ex/0106049}}.
\bibitem{k2k}
 M.~H.~Ahn {\it et al.}  [K2K Collaboration],
  Phys.\ Rev.\  D {\bf 74}, 072003 (2006)
\bibitem{min}
  N.~Tagg  [MINOS Collaboration],
{\it In the Proceedings of 4th Flavor Physics and CP Violation Conference (FPCP 2006), Vancouver, British Columbia, Canada, 9-12 Apr 2006, pp
019}
  [arXiv:hep-ex/0605058].
\bibitem{t2k}
  T.~Le,
  arXiv:0910.4211 [hep-ex].
\bibitem{nova}
  G.~J.~Feldman  [NOvA Collaboration],
{\it  In *Thomas, J.A. (ed.) et al.: Neutrino oscillations* 217-231}
\bibitem{cam57}
G.~L. Fogli and E.~Lisi, ``Tests of three-flavor mixing in long-baseline
  neutrino oscillation experiments'' {\em Phys. Rev.} {\bf D54} (1996)
  3667--3670
\bibitem{cam58}
  H.~Minakata and H.~Nunokawa,
  JHEP {\bf 0110}, 001 (2001)
  [arXiv:hep-ph/0108085].
\bibitem{cam59}
  V.~Barger, D.~Marfatia and K.~Whisnant,
  Phys.\ Rev.\  D {\bf 65}, 073023 (2002)
  [arXiv:hep-ph/0112119].
\bibitem{cam60}
  J.~Burguet-Castell, M.~B.~Gavela, J.~J.~Gomez-Cadenas, P.~Hernandez and O.~Mena,
  Nucl.\ Phys.\  B {\bf 608}, 301 (2001)
  [arXiv:hep-ph/0103258].
\bibitem{uno}
  V.~Barger {\it et al.},
  arXiv:0705.4396 [hep-ph].
\bibitem{hk}
  K.~Nakamura,
  Int.\ J.\ Mod.\ Phys.\  A {\bf 18}, 4053 (2003).
\bibitem{mem}
  A.~de Bellefon {\it et al.},
  arXiv:hep-ex/0607026.
\bibitem{ino}
  V.~M.~Datar  [INO Collaboration],
  J.\ Phys.\ Conf.\ Ser.\  {\bf 136}, 022016 (2008).
\bibitem{vhe}
  R.~Gandhi and S.~Panda,
  JCAP {\bf 0607}, 011 (2006)
  [arXiv:hep-ph/0512179].
\bibitem{gla}
  A.~Rubbia,
  J.\ Phys.\ Conf.\ Ser.\  {\bf 171}, 012020 (2009)
  [arXiv:0908.1286 [hep-ph]].
\bibitem{rg1}
  R.~Gandhi, P.~Ghoshal, S.~Goswami, P.~Mehta and S.~Uma Sankar,
  Phys.\ Rev.\ Lett.\  {\bf 94}, 051801 (2005)
  [arXiv:hep-ph/0408361]
\bibitem{rg2}
  R.~Gandhi, P.~Ghoshal, S.~Goswami, P.~Mehta and S.~Uma Sankar,
  Phys.\ Rev.\  D {\bf 73}, 053001 (2006)
  [arXiv:hep-ph/0411252].
\bibitem{rg3}
  R.~Gandhi, P.~Ghoshal, S.~Goswami, P.~Mehta, S.~U.~Sankar and S.~Shalgar,
  Phys.\ Rev.\  D {\bf 76}, 073012 (2007)
  [arXiv:0707.1723 [hep-ph]].

\bibitem{dch}
  I.~Gil-Botella  [Double Chooz Collaboration],
  J.\ Phys.\ Conf.\ Ser.\  {\bf 171}, 012067 (2009).
\bibitem{day}
  W.~Wang and f.~t.~D.~collaboration,
  arXiv:0910.4605 [hep-ex].
\bibitem{aut37}
  J.~E.~Campagne, M.~Maltoni, M.~Mezzetto and T.~Schwetz,
  JHEP {\bf 0704}, 003 (2007)
  [arXiv:hep-ph/0603172].
\bibitem{per}
  O.~L.~G.~Peres and A.~Y.~Smirnov,
  Phys.\ Lett.\  B {\bf 456}, 204 (1999)
  [arXiv:hep-ph/9902312].
\bibitem{san}
  S.~Choubey and P.~Roy,
  Phys.\ Rev.\  D {\bf 73}, 013006 (2006)
  [arXiv:hep-ph/0509197].
\bibitem{hms}
  P.~Huber, M.~Maltoni and T.~Schwetz,
  Phys.\ Rev.\  D {\bf 71}, 053006 (2005)
  [arXiv:hep-ph/0501037].
\bibitem{cam}
  J.~E.~Campagne, M.~Maltoni, M.~Mezzetto and T.~Schwetz,
  JHEP {\bf 0704}, 003 (2007)
  [arXiv:hep-ph/0603172].





















\end{thebibliography}
\end{document}